\documentclass[twocolumn,showpacs,amsmath,amssymb]{revtex4}\usepackage{graphicx}
\begin{document}
\DeclareGraphicsExtensions{.eps, .jpg}
%%%%%%%%%%%%%%%%%%%%%%%%%%%%%%%%%%%%%%%%%%
\bibliographystyle{prsty}
\input epsf

\title {Low-energy electrodynamics of superconducting diamond}
\author{M. Ortolani$^{1,2}$, S. Lupi$^{1}$, L. Baldassarre$^{1}$, U. Schade$^{2}$, P. Calvani$^{1}$, Y. Takano$^{3}$, M. Nagao$^{3}$, T. Takenouchi$^{4}$, and H. Kawarada$^{4}$}
\affiliation{$^{1}$CNR-INFM Coherentia and Dipartimento di Fisica, Universit\`a di Roma La Sapienza, Piazzale A. Moro, 2, 00185 Roma, Italy}\
\affiliation{$^{2}$Berliner Elektronenspeicherring-Gesellshaft f\"ur Synchrotronstrahlung m.b.H., Albert-Einstein Strasse 15, D-12489 Berlin, Germany}\
\affiliation{$^{3}$National Institute for Materials Science, 1-2-1 Sengen, Tsukuba 305-0047, Japan}\
\affiliation{$^{4}$School of Science and Engineering, Waseda University, 3-4-1 Okubo, Shinjuku, Tokyo 169-8555, Japan.}\

\date{\today}

\begin{abstract}
Boron-doped diamond films can become superconducting with critical temperatures $T_c$ well above 4 K. Here we first measure the reflectivity of such a film down to 5 cm$^{-1}$, by also using Coherent Synchrotron Radiation. We thus determine the optical gap 2$\Delta$, the field penetration depth $\lambda$, the range of action of the Ferrell-Glover-Tinkham sum rule, and the electron-phonon spectral function $\alpha^2F(\omega)$. We conclude that diamond behaves as a "dirty" BCS superconductor. 
\end{abstract}

\maketitle

Diamond, with its extraordinary mechanical properties, excellent thermal conductivity, and large gap between the valence and the conduction band, is potentially a semiconductor more attractive than silicon for many applications.  Therefore the transport properties of diamond films, deposited by Chemical Vapor Deposition (CVD), and doped by acceptors or donors, are being extensively explored in view of a possible, future diamond-based electronics. In this framework it has been discovered recently that heavily boron-doped diamond can also become a superconductor \cite{ekimov} below critical temperatures $T_c$ well above the liquid helium temperature \cite{takano}, if the doping level is $\agt 2.5\%$. 

Strongly covalent bonds, high concentration of impurities, and high phonon frequencies make B-doped diamond much different from the conventional metals where the Bardeen-Cooper-Schrieffer (BCS) \cite{BCS} theory of superconductivity holds.
Indeed, the metallic properties of heavily B-doped diamond are now the subject of an intense theoretical investigation. If many authors suggest that B-doped diamond in the doping regime above $\sim 0.5\%$ should be a degenerate metal \cite{lee2,fontaine}, with a conduction band strongly broadened by disorder, others point out that the deep 0.37 eV level of the isolated B-acceptor \cite{glezener} may prevent the merging of the B-like bands with the C valence band, and  propose unconventional models for the metallization of diamond  \cite{yu}. One thus may wonder whether diamond is anyhow a BCS material, eventually with a high degree of disorder, or an exotic superconductor like most of those discovered in the last two decades. The study of the electron-phonon interaction in metallic diamond, a likely candidate for the Cooper pairing mechanism, has also attracted considerable attention, since the high phonon frequencies make the adiabatic limit questionable and the covalent bonds may produce a very strong coupling costant, like in MgB$_2$ \cite{lee,boeri}.
Here we approach both this problems by first measuring the reflectivity of a superconducting diamond film, in the sub-Terahertz region down to 5 cm$^{-1}$ where the gaps of superconductors are observed, and in the infrared region, where the signatures of the electron-phonon coupling appear. The sub-Terahertz frequencies have been reached, with the required signal-to-noise ratio, by use of Coherent Synchrotron Radiation.

A basic feature of the superconducting state is the opening, for $T < T_c$, of a gap $E_g$ in the electronic density of states. Correspondingly, if the Cooper pairs are in a spherically symmetric $s$ state, the reflectivity becomes $R_s(\omega) = 1$ for any $\omega \leq  2\Delta(T)$, the optical gap, where $hc\Delta \sim E_g$. Above $T_c$ and in the same low-frequency range, the reflectivity of the normal metal is instead described by the Hagen-Rubens formula $R_n(\omega) = 1 - [8\omega\Gamma(T)/{\omega_p}^2]^{\frac{1}{2}}$, where $\Gamma(T)$ is the relaxation rate of the carriers and  $\omega_p$ their plasma frequency. Therefore, if the metal is in the "dirty" regime defined by $\Gamma(T_c) > 2\Delta(0)$, the ratio $R_s/R_n$ exhibits a peak at $2\Delta$.  This property allows one to measure the optical gap \cite{basov}. Early studies indicate that boron-doped diamond films are in the dirty limit and display a highly symmetric wave function \cite{jap_nature}.  An infrared determination of their optical gap is then possible, and the result can be compared with the BCS prediction $2hc\Delta(0)/k_BT_c = 3.52$. Further information may be provided by the optical conductivity $\sigma(\omega)$ that one extracts from the raw reflectivity data. By comparing its values in the gap region measured below and above $T_c$, and applying suitable sum rules, one can obtain the field penetration depth $\lambda$. In the clean limit defined by $\Gamma(T_c) \ll 2\Delta(0)$ $\lambda$ coincides with the London penetration depth $\lambda_L$, while in the dirty limit $\lambda\sim\lambda_L(\Gamma/\Delta)^\frac{1}{2}$ \cite{DresselGruner}. In the present experiment we have obtained both $2\Delta$ and $\lambda$, in addition to other relevant parameters discussed below, from the reflectivity of superconducting diamond measured at different temperatures from 5 to 20000 cm$^{-1}$. 

\begin{figure}[tbp]
  %\begin{center}
   %\resizebox{9.0 cm}{!}{\includegraphics{Figura1.epsf}}
    \leavevmode
    \epsfxsize=8cm \epsfbox {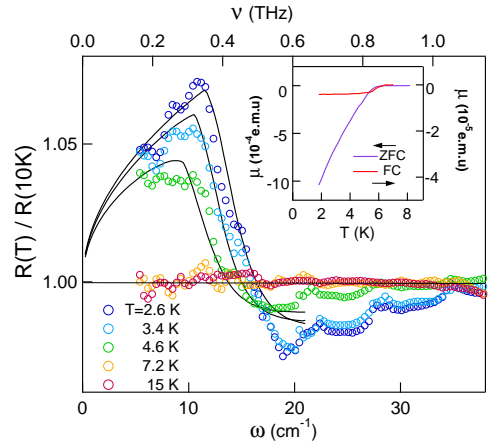}
     \caption{Reflectivity  of strongly boron-doped diamond at different temperatures in the sub-THz region, normalized to its values at $10$ K. The peak frequency yields $2\Delta$ at that temperature. The lines are fits obtained by assuming a BCS reflectivity below $T_c$ and a Hagen-Rubens model at $10$ K. The inset shows the magnetic moment of the sample, as cooled either in a $10$ Oe field (red line, FC) or in zero field (violet line, ZFC). }
\label{delta}
%\end{center}
\end{figure}

The sample was a film about $3\mu$m thick, $2.5x2.5$ mm wide, grown by CVD and deposited on pure CVD diamond \cite{takano}. The boron concentration was estimated to be $\sim 6x10^{21}$ cm$^{-3}$. The sample magnetic moment $\mu(T)$ is reported in the inset of Fig. \ref{delta}. It shows the superconducting transition with an onset at $T_c = 6$ K. Such a low value of $T_c$ implies a BCS optical gap of the order of 10 cm$^{-1}$, hardly accessible to standard infrared sources. In order to detect the small difference between $R_s$ and $R_n$ at such sub-THz frequencies (1 THz = 33 cm$^{-1}$), the sample was illuminated by Coherent Synchrotron Radiation (CSR) extracted from the electron storage ring BESSY, working in the so-called low-$\alpha$ mode with a beam current $i \sim 20$ mA \cite{Abo}. CSR is free of thermal noise and, in the sub-THz range, is more brilliant than any other broad-band radiation source by two orders of magnitude at least \cite{Abo}. The thickness of the film was such that no correction for multiple reflections was needed, as confirmed by the absence of fringes in the reflectivity spectra. Similarly, rigorous checks excluded any effect of diffraction, under the measuring procedure described below. By using CSR, a commercial interferometer and a bolometer working at $1.6$ K we obtained, in the optical gap region, an error on the reflected intensity $\Delta I_R/I_R = \pm$ 0.3 \%. From 20 to 40 cm$^{-1}$ we use ordinary synchrotron radiation from the same bending magnet with the ring working at $\sim$ 200 mA. Using an automatic, remotely driven mirror system, the dependence of $I_R$ on the slowly-decaying electron-beam current in the ring was exactly taken into account by continuously measuring both the intensity reflected by the sample $I_R(\omega ,T, i)$ and the intensity transmitted trough the interferometer $I_0(\omega, i)$. The reliability of the above procedure was tested by cycling the sample temperature from 30 to 2.6 K and back several times, for several beam current values. For 40 $<\omega <20000$ cm${-1}$ we used conventional sources. 

The ratio $I_R(T)/I_R(10 K) = R_s(T)/R_n(10 K)$ is reported in Fig. \ref{delta}. The three curves at $T < T_c$ exhibit a strong frequency dependence in the sub-THz region, and peak at a frequency roughly corresponding to the optical gap $2\Delta(T)$. As a cross-check, the $I_R(T)/I_R(10 K)$ data taken at $T > T_c$ do not show any peak and equal 1 within the noise. The peak value can be straightforwardly compared with $T_c$ to check the BCS prediction $2E_g/k_BT_c = 3.52$. From a first inspection of our sub-Terahertz data, we find at $T=2.6 $ K a peak value $\simeq 12 $ cm$^{-1}$ and then $2hc\Delta(2.6 $K$)/k_BT_c \simeq 3$. This finding motivates us to use a BCS framework to fit the data.
The good fits reported in Fig. \ref{delta} were obtained by modeling the complex conductivity $\sigma(\omega)$ in the normal and in the superconducting state separately to obtain a calculated ratio $R(\omega ,T)/R(\omega ,10$ K$)$. In the normal state we used the conventional Drude model with $\omega_p$ and  $\Gamma$ as free parameters, while below $T_c$ we used the energy-integrated Green function method of Zimmermann et al. \cite{zimmer} with $\sigma_{dc}={\omega_p}^2/\Gamma$ and $\Delta$ as free parameters and a fixed  $T_c = 6$ K. The fit gave $\sigma_{dc}(T_c) = 340 \pm$ 40 $\Omega^{-1}$ cm$^{-1}$, in good agreement with dc transport measurements on B-doped diamond films with similar doping and $T_c$ \cite{takano}. Furthermore, the lineshape seems to be well descibed by the BCS curve. The main output of the fit, however, is the gap value, which at 4.6, 3.4, and 2.6 K is found to be $2\Delta$ = 9.5, 10.5, and 11.5 cm$^{-1}$, respectively. This leads to an extrapolated value \cite{DresselGruner} $2\Delta(0)$ = 12.5 cm$^{-1}$, or $2hc\Delta(0)/k_BT_c = 3.0\pm 0.5$, in satisfactory agreement with the above BCS prediction.

\begin{figure}[tbp]
   %\begin{center}
   %\resizebox{9.0 cm}{!}{\includegraphics{Figura1.epsf}}
    %\leavevmode
    \epsfxsize=8cm \epsfbox {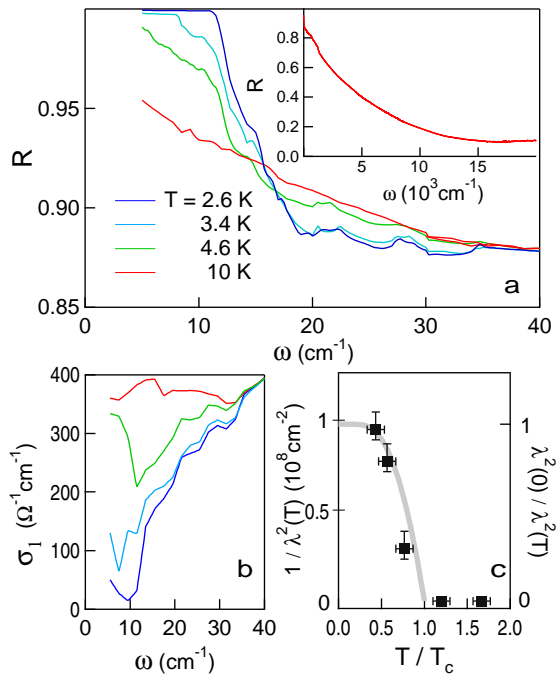}
     \caption{Optical response of superconducting diamond: in a), the reflectivity of the film as it comes out by multiplying the ratios of Fig.\ref{delta} by the absolute $R(\omega)$ at $10$ K, also shown in the inset in the full measuring range;  in b), the real part of the optical conductivity which shows gap opening and full recovery for $\omega > 6\Delta$, as expected for a BCS superconductor in the dirty limit; in c), the inverse square of the penetration depth, obtained from missing area in b) (experimental points), is compared with its behavior for a dirty BCS superconductor, once normalized to zero temperature (grey line).}
\label{due}
%\end{center}
\end{figure}

Afterwards, we obtained the absolute reflectivity $R_n(10 K)$ up to 20000 cm$^{-1}$ (inset of Fig. \ref{due}a) by extending the measuring range as reported above, and taking as reference the film itself, coated with a gold or silver layer evaporated \textit{in situ}.
The reflectivity in the superconducting phase was then reconstructed as $R_s(T) = [I_R(T)/I_R(10 K)] R_n(10 K)$ (see Fig. \ref{due}a), and used to obtain the optical conductivity   $\sigma(\omega) = \sigma_1(\omega)+i\sigma_2(\omega)$ by standard Kramers-Kronig transformations. The real, or absorptive part  $\sigma_1(\omega)$, reported in Fig. \ref{due}b, decreases in the sub-THz range for $T < T_c$, due to the opening of the optical gap. At $4.6$ K, a residual quasi-particle contribution can still be distinguished at the lowest measured frequencies. At $2.6$ K, zero absorption is attained below $\sim 10$ cm$^{-1}$, a value comparable to $2\Delta$ obtained from the reflectivity fitting. One may notice that the optical conductivity of the superconducting phase  ${\sigma_1}^s(\omega)$ and that of the normal phase ${\sigma_1}^n(\omega)$ in Fig. \ref{due}b coincide for $\omega \agt 35$ cm$^{-1} \sim 6\Delta$, indicating a dirty limit behavior of a BCS superconductor  \cite{MattisBard}. According to the Ferrell-Glover-Tinkham sum rule \cite{MattisBard} the area $A$ removed at $T < T_c$ below ${\sigma_1}(\omega,T)$, builds up the collective mode at $\omega = 0$. The spectral weight condensed into this peak, 

\begin{equation}
A = \int^{6\Delta}_0({\sigma_1}^s-{\sigma_1}^n)d\omega 
\label{spectralweight}
\end{equation}

\noindent may be used to extract the penetration depth $\lambda= 1/2 \pi(8A)^\frac{1}{2}$ \cite{DresselGruner}. We thus find out $\lambda \simeq 1 \mu$m at 2.6 K. As already mentioned, in the dirty limit $\lambda\sim\lambda_L(\Gamma/\Delta)^frac{1}{2}$. One may thus estimate $\lambda\sim 50$ nm for our diamond film at 2.6 K. The large difference between $\lambda$ and $\lambda_L$ is usually related to a large impurity scattering in the presence of disorder, driving the superconductor to the dirty limit. Finally, $1/\lambda^2$ is plotted in Fig.\ref{due}c vs. $T/T_c$, and compared with the BCS prediction for a dirty superconductor \cite{DresselGruner}. Once again, this model appears to well describe the behavior here observed in boron-doped diamond.  

\begin{figure}[tbp]
   %\begin{center}
   %\resizebox{9.0 cm}{!}{\includegraphics{Figura1.epsf}}
    %\leavevmode
    \epsfxsize=8cm \epsfbox {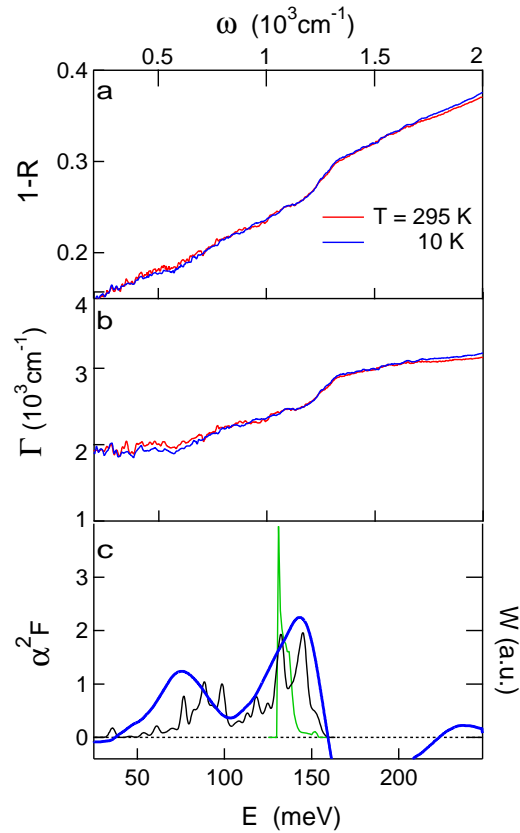}
     \caption{Determination of the electron-phonon spectral function in hole-doped diamond. \textbf{a.} The raw absorbance $1-R(\omega)$ at two temperatures. \textbf{b.} The frequency-dependent relaxation rate of the carriers, as obtained at those temperatures from the conductivity $\sigma(\omega)$ by use of Eq. \ref{gamma}. \textbf{c.} The blue line is the function $W(\omega)$ defined in Eq. \ref{vdoppio}, as calculated from the $\Gamma$ values in \textbf{b}. The green and the black lines are the electron-lattice spectral functions $\alpha^2F(\omega)$ reported for diamond at \(3\%\) boron doping in Ref. \cite{boeri} and \cite{xiang}, respectively.}
\label{tre}
%\end{center}
\end{figure}

Since the BCS theory seems to describe well the electrodynamics of the superconducting state of diamond, it is worth to analyze the optical response in the far- and mid-infrared (100 to 2000 cm$^{-1}$) to obtain information on the electron-phonon interaction. Indeed, optical phonon modes interacting with the charges are likely to be responsible for the Cooper pairing in a BCS framework. These could be optical zone-center phonons coupled to holes at the top of the diamond valence band in a degenerate metal scenario \cite{lee,boeri}, as well as modes generated by boron atoms, which may couple to holes within a larger momentum distribution \cite{xiang,blase}.
The electron-phonon spectral function $\alpha^2F(\omega)$, also named the optical Eliashberg function \cite{marsiglio}, displays characteristic resonances at the frequencies of the phonons interacting with the carriers. In turn, $\alpha^2F(\omega)$ can be extracted from the dielectric function $\epsilon(\omega) = \epsilon _1(\omega)+i\epsilon_2(\omega)$, which is obtained from the Kramers-Kronig transformations of $R(\omega)$. As in the presence of electron-lattice interactions the carrier scattering rate $\Gamma(T)$ may also become a function of $\omega$, one first determines \cite{timusk} 

\begin{equation}
\Gamma(\omega)= \frac{(\omega_p^2/\omega)\epsilon_2(\omega)}{\{[\epsilon_1(\omega)-\epsilon_{\infty}]^2 + \epsilon_2^2(\omega)\}^2} .
\label{gamma}
\end{equation}                                                                     

Here $\omega_p$ is the plasma frequency of the normal carriers, and $\epsilon_{\infty}$ is the high-frequency contribution to the dielectric function. Then one should solve the integral equation \cite{timusk}

\begin{equation}
\Gamma(\omega)= \frac{2\pi}{\omega} \int^{\omega}_{0} (\omega-\Omega)\alpha^2F(\omega)d\Omega\label{gamma2}
\end{equation}                               

\noindent which holds for $T = 0$. However, the solution of Eq. \ref{gamma2} is not always unambiguous \cite{dordevic}. Alternatively, one may obtain a reliable estimate of $\alpha^2F(\omega)$ in the phonon region \cite{marsiglio} by double-differentiating $\Gamma(\omega)$ extracted from the dielectric function through Eq. \ref{gamma} and calculating the quantity                                      

\begin{equation}
W(\omega)= \frac{1}{2\pi}\frac{d^2}{d\omega^2}[\omega\Gamma(\omega)]
\label{vdoppio}
\end{equation}                                                                            

\noindent whose frequency dependence is closely related to that of $\alpha^2F(\omega)$. The raw absorbance $1-R(\omega)$ of the boron-doped film in the far- and mid-infrared is shown in Fig. \ref{tre}a and the corresponding $\Gamma(\omega)$ in Fig. \ref{tre}b. Both quantities exhibit a clear deviation from a quasi-linear behavior vs. $\omega$ around 1200 cm$^{-1}$. A change of slope in $\Gamma(\omega)$ is also detected at  about 500 cm$^{-1}$. Both those features reflect in the $W(\omega)$ shown in Fig. \ref{tre}c and obtained from Eq.\ref{vdoppio}. Therein, the results are reported in arbitrary units, as their absolute values are sensitive to the smoothing procedure needed to differentiate $\Gamma(\omega)$, the extent of which is somewhat arbitrary. However, the spectral shape of $W(\omega)$ is robust against that procedure. In order to check our result, in addition to using Eq.\ref{vdoppio} we solved Eq.\ref{gamma2} by the numerical method reported in Ref.\cite{dordevic}. We thus obtained a family of curves, depending on the number of accepted poles \cite{dordevic}, similar in shape to that plotted in Fig. \ref{tre}c. However, as reported in the literature for other systems, here also some unphysical negative values of $\alpha^2F(\omega)$ could not be eliminated.

The shape of $W(\omega)$ is compared in Fig. \ref{tre}c with two calculated  functions  $\alpha^2F(\omega)$ recently reported \cite{boeri,xiang} for diamond with 3\%  boron impurities, as it is the case of our sample. Both theoretical predictions, where most of the electron-phonon interaction is provided by the optical phonon branch around 1200 cm$^{-1}$, are confirmed by our experimental estimate of  $\alpha^2F(\omega)$. The agreement between theory and experiment is particularly impressive for the calculation of Ref. \cite{xiang}, where the distortion of boron impurities on the diamond lattice is fully taken into account. As a result, new modes appear around $600$ cm$^{-1}$, which according to the calculations and to the present data are also involved in the charge-lattice interaction. One may notice that the Migdal-Eliashberg  approximation $\omega_{ph} \ll E_F/hc \simeq 4800$ cm$^{-1}$ \cite{lee}, where $E_F$ is the Fermi energy of diamond, is more suitable for those modes than for the main optical branch  around $1200$ cm$^{-1}$.

In conclusion, we have studied here the electrodynamics of strongly boron-doped, superconducting diamond, down to the sub-THz region. Therein we have exploited Coherent Synchrotron Radiation, which allows one to improve the signal-to-noise ratio by orders of magnitude using a conventional Michelson interferometer. We have thus clearly observed the opening of an optical gap in hole-doped diamond below $T_c$, whose width 2$\Delta$ is $3k_BT_c$ at $T = 0$, and verified that the Ferrell-Glover-Tinkham sum rule holds within approximately 6$\Delta$. The field penetration depth is found to be 1 $\mu$m at 2.6 K. Finally, we have determined the electron-phonon spectral function $\alpha^2F(\omega)$, which can greatly help to identify the mediators of Cooper pairing. In agreement with recent calculations, our data show that the charge-lattice interaction involves mainly the optical phonon branch of pure diamond, but also additional modes at lower frequencies, induced by doping. The ensemble of these results builds up a consistent picture of superconducting diamond, where it behaves as a BCS superconductor in the dirty limit.

\end{document}